\documentclass[%
 reprint,
 superscriptaddress,
 amsmath,amssymb,
 aps,
prb,
]{revtex4-2}

\usepackage{dcolumn}
\usepackage{bm}
\usepackage{amsmath}
\usepackage{txfonts}
\usepackage[T1]{fontenc}
\usepackage{xspace}
\usepackage{ulem}
\usepackage{comment}
\usepackage{bbold}
\usepackage{braket}
\usepackage{ascmac}

\ifx\pdfoutput\undefined
\usepackage[dvipdfmx]{graphicx}
\usepackage[dvipdfmx]{hyperref}
\usepackage[dvipdfmx]{color}
\usepackage[dvipdfmx]{xcolor}
\else
\usepackage{graphicx}
\usepackage{hyperref}
\usepackage{color}
\usepackage{xcolor}
\fi

\DeclareRobustCommand{\DEL}{\bgroup\markoverwith{\textcolor{red}{\rule[.5ex]{2pt}{1.2pt}}}\ULon}

\newcommand{\hstv}{{\bf h}^{\rm stat}}
\newcommand{\hacv}{{\bf h}^{\rm AC}}
\newcommand{\hst}{h^{\rm stat}}

\newcommand{\hatv}[1]{\hat{\bf #1}}

\newcommand{\chimax}[1]{\chi_{E, #1}^{\rm max}}
\newcommand{\ImchiS}[1]{{\rm Im}\,\chi_{S, #1}(\omega)}

\graphicspath{
{./}
}

\begin{document}

%Title of paper
\title{
Emergent electric field from magnetic resonances in a one-dimensional chiral magnet
}

\author{Kotaro Shimizu}
\author{Shun Okumura} 
\author{Yasuyuki Kato}
\author{Yukitoshi Motome}
\affiliation{Department of Applied Physics, The University of Tokyo, Tokyo 113-8656, Japan}

\date{\today}

\begin{abstract}
The emergent electric field (EEF) is a fictitious electric field acting on conduction electrons through the Berry phase mechanism. 
The EEF is generated by the dynamics of noncollinear spin configurations and becomes nonzero even in one dimension. 
Although the EEF has been studied for several one-dimensional chiral magnets, 
most of the theoretical studies were performed in limited situations with respect to the strength and direction of the magnetic fields.
Furthermore, the effect of edges of the system has not been clarified, whereas it can be crucial in nano- and micro-scale samples.
Here, we perform a comprehensive theoretical study on the momentum-frequency profile of the EEF in a one-dimensional chiral magnet using the Landau-Lifshitz-Gilbert equation while changing the strength and direction of the static and AC magnetic fields for both bulk and finite-size chains with edges. 
From the bulk calculations under the periodic boundary condition, we find that the EEF is resonantly enhanced at the magnetic resonance frequencies; 
interestingly, the higher resonance modes are more clearly visible in the frequency profile of the EEF response than in the magnetic one. 
Furthermore, we show that the EEF is amplified along with the solitonic feature of the spin texture introduced by the static magnetic field perpendicular to the chiral axis. 
We also show that the static magnetic field parallel to the chiral axis drives the EEF in the field direction, 
in addition to much slower drift motion in the opposite direction associated with the Archimedean screw dynamics, suggesting a DC electric current generation. 
For the finite-size chains under the open boundary condition, we find additional resonance modes localized at the edges of the system that are also more clearly visible in the EEF response than the magnetic one. 
Moreover, we show that a substantial EEF is generated from the edges even in the fully-polarized forced-ferromagnetic phase, although it is absent in the bulk case. 
Our results reveal that the emergent electric phenomena in one-dimensional chiral magnets can be tuned by the magnetic field and the sample size, and provide not only a good probe of the magnetic resonances 
but also a platform for the applications to electronic devices.

\end{abstract}

%\keywords{}

\maketitle

\section{Introduction \label{sec:1}}

The Berry phase --- a phase factor acquired in an adiabatic motion of quantum particles~\cite{Berry1984} brings about a fictitious electromagnetic field~\cite{Sundaram1999,Xiao2010}, which leads to intriguing quantum transport and optical phenomena for electrons in solids. 
For instance, the fictitious magnetic and electric fields arising from the Berry phase in momentum space bring about a quantum Hall effect~\cite{Thouless1982} and a quantized charge pumping~\cite{Thouless1983}, respectively. 
Meanwhile, the Berry phase in real space is also of importance, especially in magnets, where
the fictitious electromagnetic fields, often called the emergent electromagnetic fields, are generated by noncollinear and noncoplanar spin textures~\cite{Volovik1987,Xiao2010,Nagaosa2012-1,Nagaosa2012-2,Nagaosa2013}. 
The emergent magnetic field arises as a fictitious magnetic flux through a plaquette when the surrounding spins are noncoplanar, i.e., when the scalar spin chirality is nonzero, and hence, it requires two- or three-dimensional noncoplanar spin textures.  
The typical example is found in magnetic skyrmions~\cite{Bogdanov1989, Bogdanov1994, Bogdanov1995, Roessler2006}, which give rise to the topological Hall effect via the emergent magnetic field~\cite{Loss1992, Ye1999, Bruno2004, Onoda2004, Binz2008}. 
In contrast, the emergent electric field (EEF) does not require noncoplanar spin configurations; it arises from time evolution of noncollinear spin textures, and hence, can be generated even in one-dimensional systems~\cite{Stern1992, Barnes2007, Korenman1977, Berger1986, Volovik1987}. 
Therefore, the emergent electric phenomena arise in a much wider range of magnetic materials 
than the emergent magnetic phenomena.

Quantum phenomena arising from the EEF in magnets have been intensively studied over the past decade. 
Experimental detection of the EEF has been done for field-induced motion of a magnetic domain wall~\cite{Yang2009}, ferromagnetic resonance in a patterned ferromagnetic film~\cite{Yamane2011}, and gyrating motion of a magnetic vortex~\cite{Tanabe2012}. 
Recently, the EEF has attracted renewed interest in the dynamics induced by an electric current. 
It was pointed out that the EEF by a current-induced motion of swirling spin textures gives rise to an inductance~\cite{Nagaosa2019}, and such behavior was observed in helical magnets~\cite{Yokouchi2020} even at room temperature~\cite{Kitaori2021emergent}.

The central question that we address in this study is how to enhance the EEF in magnets by an external magnetic field. 
To clarify the fundamental behavior of the EEF, we will focus on a simple model for one-dimensional chiral magnets. 
Thus far, the EEF in one-dimensional chiral magnets has been theoretically studied, mostly 
by using the collective coordinate method in continuum space~\cite{Kishine2012,Ovchinnikov2013,Kishine2016}.
However, such analytical studies were limited with respect to the strength and direction of the magnetic fields and comprehensive studies have not yet to be done, even though the system shows a variety of spin textures and their magnetic excitations depending on the fields. 
In addition, as the EEF is generated by the dynamics of spin textures, it is important to elucidate the relationship between the EEF and the magnetic excitations, but a large part of it also remains unexplored. 
To clarify these issues, it is desired to systematically study the dependence of the EEF on the static and dynamical magnetic fields beyond the previous studies. 

In addition, we aim at investigating the contributions to the EEF from not only bulk but also edges of the system, since the latter can be crucial in experiments for nano- and micro-scale samples. 
Effect of edges has been discussed for both static and dynamical properties in magnets. 
For instance, in one-dimensional chiral magnets, the magnetization varies discretely as a function of the magnetic field in finite-size systems with edges~\cite{Wilson2013, Kishine2014, Togawa2015, Wang2017, Shinozaki2018}. 
It was also shown that the edges affect the magnetic resonance and generate additional edge modes~\cite{Hoshi2020}. 
Given the edge effects on the spin textures and the magnetic excitations, the EEF is expected to be strongly influenced by the presence of edges as well, but the systematic study has not yet been conducted.

In this paper, we perform a systematic theoretical study of the EEF in a one-dimensional chiral magnet.  
By numerical simulations of the real-space and real-time spin dynamics based on the Landau-Lifshitz-Gilbert (LLG) equation, we clarify the momentum-frequency profile of the EEF while changing the strength and direction of the static and AC magnetic fields for the systems under the periodic boundary condition (PBC) and the open boundary condition (OBC). 
For the bulk response calculated under the PBC, we show that the EEF becomes prominent at the magnetic resonance frequencies and it is enlarged along with the solitonic feature enhanced by the perpendicular component of the 
static magnetic field to the chiral axis, regardless of the direction of the AC magnetic field. 
We find that the higher frequency resonance modes are more clearly visible in the frequency profile of the EEF response than the magnetic one. 
Furthermore, by analyzing the spatiotemporal profiles, we also show that the parallel component of the static magnetic field drives the EEF in the field direction due to the magnon propagation, in addition to much slower drift to the opposite direction due to the Archimedean screw dynamics~\cite{Nina2021}. 
Meanwhile, for the edge contributions calculated under the OBC, we find additional resonance modes of the EEF localized at the system edges, whose response can be comparable to or larger than the bulk ones, depending on the system size, and again more visible than the magnetic response.

In addition, we show that the EEF is generated even in the fully-polarized forced-ferromagnetic (FFM) phase in the presence of the edges. 
These systematic analyses provide comprehensive understanding of the EEF associated with the magnetic resonance in the one-dimensional chiral magnet, which would be useful for further exploration of emergent electric phenomena.

The rest of the paper is organized as follows. 
In Sec.~\ref{sec:2}, we introduce the model (Sec.~\ref{sec:model}), the numerical method (Sec.~\ref{sec:LLGeq}), and the definitions of the physical quantities (Sec.~\ref{sec:quantity}) used in the following analyses. 
In Sec.~\ref{sec:PBC}, we show the results for the bulk contribution obtained under the PBC. 
We first show the complex admittance of the EEF for various combinations of the static and AC magnetic fields (Sec.~\ref{sec:PBC.1}), and then, compare the results with the dynamical spin susceptibility (Sec.~\ref{sec:PBC.2}). We also show the spatiotemporal profiles of the spin textures and the EEF in the resonance modes (Sec.~\ref{sec:PBC.3}). 
In Sec.~\ref{sec:OBC}, we show the results for the edge contribution obtained under the OBC in the similar manner in Sec.~\ref{sec:PBC}. 
We discuss the results in Sec.~\ref{sec:discussion}. 
Section~\ref{sec:summary} is devoted to the summary of this paper.

%%%===========================================================================
%%%===========================================================================
%%% sec:2
%%%===========================================================================
%%%===========================================================================
\section{Model and method \label{sec:2}}

In this section, we introduce the model for a one-dimensional chiral magnet and the numerical method to study the dynamics of the model. 
In Sec.~\ref{sec:model}, we introduce the model Hamiltonian with the setting of parameters and two types of boundary conditions. 
We describe the method based on the LLG equation in Sec.~\ref{sec:LLGeq} %, 
and the definitions of the quantities to be measured in Sec.~\ref{sec:quantity}.

%%%---------------------------------------------------------------------------
%%% fig:setup_schematic
\begin{figure}[tb]
\centering
\includegraphics[width=1.0\columnwidth]{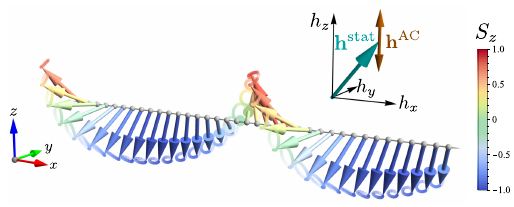}
\caption{
\label{fig:setup_schematic}
Schematic picture of the setup in this study. 
The arrows represent spin configurations in the model in Eq.~\eqref{eq:ham}; 
the color denotes the $S_z$ component of spins. 
The pale curves attached to the arrowheads represent the trajectories of the spins in the time evolution under the static magnetic field ${\bf h}^{\rm stat}$ and the AC magnetic field ${\bf h}^{\rm AC}$ depicted in the inset.
}
\end{figure}
%%%---------------------------------------------------------------------------

%%%===========================================================================
%%% sec:model

\subsection{Model \label{sec:model}}
In this study, we consider a one-dimensional chiral magnet, which is described by the time-dependent Hamiltonian given by 
\begin{eqnarray}
\mathcal{H}(t)=
\sum_l && \bigl[ -J{\bf S}_l(t)\cdot{\bf S}_{l+1}(t) 
-D\hat{\bf x} \cdot \left( {\bf S}_l(t) \times {\bf S}_{l+1}(t) \right) \notag \\ 
&& ~+ {\bf h}(t)\cdot{\bf S}_l(t) \bigr], 
\label{eq:ham}
\end{eqnarray}
where $t$ is time, and ${\bf S}_l(t)$ represents the classical spin at site $l$ and time $t$ with $|{\bf S}_l(t)|=1$. 
The first and second terms in the square brackets denote the Heisenberg and Dzyaloshinskii-Moriya (DM) interactions, respectively. 
The DM vector is taken along the chain direction parallel to the unit vector $\hat{\bf x}$. 
Hereafter, we set the energy scale as $J=1$ and the lattice constant as unity, and take $D=\tan\left(\frac{\pi}{10}\right)$. 
This stabilizes a helical spin structure in the equilibrium state at zero magnetic field, in which the spins rotate in the $yz$ plane with a period of $20$ lattice sites. 
The last term in Eq.~\eqref{eq:ham} describes the Zeeman interaction, where ${\bf h}(t)$ represents a time-dependent external magnetic field in the unit of $g\mu_{\rm B}$, where $g$ is the electron g-factor and $\mu_{\rm B}$ is the Bohr magneton. 
Note that we take the convention of a positive sign for this term. 
In the following, we consider both static and AC magnetic fields, denoted by ${\bf h}^{\rm stat}$ and ${\bf h}^{\rm AC}(t)$, respectively, as 
\begin{eqnarray}
{\bf h}(t)={\bf h}^{\rm stat}+{\bf h}^{\rm AC}(t). 
\label{eq:def_magneticfield}
\end{eqnarray}
See Fig.~\ref{fig:setup_schematic} for the setup of the model.

When only the static field is applied, the model in Eq.~\eqref{eq:ham} stabilizes a conical spin structure, a chiral soliton lattice (CSL) and their mixture, in addition to the FFM, depending on the field strength and direction~\cite{Dzyaloshinskii1964, Izyumov1984, Kishine2005, Victor2016, Masaki2018}. 
In this study, without loss of generality, we consider  the static magnetic field in the $xz$ plane, i.e., $\hstv=(\hst_x, 0, \hst_z)$. 
Then, by increasing $\hst_x$ with $\hst_z=0$, the spins on the $yz$ plane in the zero-field helical state cant uniformly in the $x$ direction to form the conical spin structure. 
In this case, the magnetic period is unchanged, until the phase transition to the FFM phase. 
In contrast, when $\hst_x=0$, $\hst_z$ introduces a solitonic feature in the spin structure, leading to the CSL. 
In this case, the magnetic period increases with $\hst_z$ and diverges at the transition to the FFM state. 
When both $\hst_x$ and $\hst_z$ are nonzero, a complicated spin texture with a mixture of the conical state and the CSL is stabilized, as exemplified in Fig.~\ref{fig:setup_schematic}
. 
We will study how the AC field modulates these spin textures and generates the EEF.

In the following calculations, to clarify not only bulk but also edge contributions, we compare the results for the systems under the PBC and the OBC. 
In both cases, we set the system size as $L=10^3$ spins, except for the study of the system size dependence in the OBC case in Sec.~\ref{sec:OBC.2}. 
For the PBC case, we confirm that $L=10^3$ is sufficiently large to study the bulk contributions. 
The sum in Eq.~(\ref{eq:ham}) is taken for $l=0, 1, \cdots, L-1$, where we impose 
${\bf S}_{L}(t)={\bf S}_0(t)$ for the PBC case, whereas 
${\bf S}_{L}(t)=0$ for the OBC case. 

%%%===========================================================================

%%%===========================================================================
%%% sec:LLGeq

\subsection{Landau-Lifshitz-Gilbert equation \label{sec:LLGeq}}

We study the real-time dynamics of the model in Eq.~(\ref{eq:ham}) by using the LLG equation given by~\cite{Landau1935,Gilbert1955} 
\begin{eqnarray}
\frac{\partial{\bf S}_l(t)}{\partial t}=\frac{1}{1+\alpha^2}&& \left[
-{\bf S}_l(t)\times{\bf h}^{\rm eff}_l(t) \right. \notag \\
&& \ \left.+\alpha {\bf S}_l(t)\times\left({\bf S}_l(t)\times{\bf h}^{\rm eff}_l(t)\right)\right], 
\label{eq:LLG}
\end{eqnarray}
where $\alpha$ is the Gilbert damping and ${\bf h}^{\rm eff}_l(t)$ is the mean magnetic field at time $t$ defined by 
\begin{eqnarray}
{\bf h}^{\rm eff}_l(t) = 
\frac{\partial \mathcal{H}(t)}{\partial {\bf S}_l(t)} 
&=& 
-J\bigl({\bf S}_{l+1}(t)+{\bf S}_{l-1}(t) \bigr) \notag \\
&&+ D\hat{\bf x}\times \bigl( {\bf S}_{l+1}(t) - {\bf S}_{l-1}(t)\bigr) + {\bf h}(t). 
\label{eq:heff}
\end{eqnarray}
Note that the length constraint of $|{\bf S}_l(t)|=1$ is deferred only when taking the derivative of the Hamiltonian with respect to ${\bf S}_l(t)$. 
Here, we impose ${\bf S}_{-1}(t)={\bf S}_{L-1}(t)$ and ${\bf S}_{-1}(t)=0$ for the PBC and OBC cases, respectively. 
We numerically solve Eq.~(\ref{eq:LLG}) by using the fourth-order Runge-Kutta method with a time step $\Delta t=0.02$.
In the following calculations, we take $\alpha=0.04$, which is a typical value for ferromagnetic metals~\cite{Mizukami2001, Mizukami2010,Oogane2006,Oogane2010}.
%%%===========================================================================

%%%===========================================================================
%%% sec:quantity

\subsection{Physical quantity \label{sec:quantity}}
The EEF, which is generated by the time evolution of noncollinear spin textures, is calculated as 
\begin{eqnarray}
\bar{E}^{\rm em}(t)=\frac{1}{L}\sum_l E^{\rm em}_l(t), 
\label{eq:eef}
\end{eqnarray}
with
\begin{eqnarray}
E^{\rm em}_l(t)={\bf S}_l(t)\cdot\left(\hat{\delta} {\bf S}_{l}(t)\times\frac{\partial {\bf S}_l(t)}{\partial t}\right), 
\label{eq:eefloc}
\end{eqnarray}
from the spin structure obtained by numerically solving the LLG equation in Eq.~(\ref{eq:LLG}).
In Eq.~(\ref{eq:eefloc}), the spatial derivative $\frac{\partial {\bf S}_l}{\partial x}$ in continuum space~\cite{Volovik1987,Xiao2010,Nagaosa2012-1,Nagaosa2012-2,Nagaosa2013} is calculated 
by the discrete difference $\hat{\delta}{\bf S}_l(t)=\frac{1}{2}\left({\bf S}_{l+1}(t)-{\bf S}_{l-1}(t)\right)$. 
For the edges in the OBC case, we use
$\hat{\delta}{\bf S}_0(t)=-\frac{3}{2}{\bf S}_0(t)+2{\bf S}_1(t)-\frac{1}{2}{\bf S}_2(t)$ and
$\hat{\delta}{\bf S}_{L-1}(t)=\frac{3}{2}{\bf S}_{L-1}(t)-2{\bf S}_{L-2}(t)+\frac{1}{2}{\bf S}_{L-3}(t)$ to reduce the discretization errors~\cite{Fornberg1988}.

We compute the complex admittance of the EEF defined as 
\begin{eqnarray}
\chi_{E,\mu}(\omega) = \frac{\bar{E}^{\rm em}(\omega)}{h_{\mu}^{\rm AC}(\omega)}, 
\label{eq:chieef}
\end{eqnarray}
where $\mu=x,y,z$ and the Fourier component of the quantity $\mathcal{O}(t)$ is obtained by 
\begin{equation}
\mathcal{O}(\omega)=\frac{1}{N_{t}}\sum_{n=0}^{N_{t}-1} \mathcal{O}(t_n)e^{-i\omega t_n}. 
\end{equation}
Here, we measure the quantity every 50 time steps, and the time at $n$th measurement and the total number of measurements are denoted by $t_n=50n\Delta t=n$ and $N_t$, respectively. 
We also calculate the dynamical spin susceptibility defined as
\begin{eqnarray}
\chi_{S, \nu\mu}(\omega)=\frac{\Delta \bar{S}_{\nu}(\omega)}{h_{\mu}^{\rm AC}(\omega)}, 
\label{eq:chimag}
\end{eqnarray}
where $\Delta\bar{\bf S}(t)=\frac{1}{L}\sum_l \left[{\bf S}_l(t)-{\bf S}_l(0)\right]$ is time variation of the magnetization.

In the actual calculations of $\chi_{E,\mu}(\omega)$ and $\chi_{S,\nu\mu}(\omega)$ in Secs.~\ref{sec:PBC} and \ref{sec:OBC}, 
we apply a pulse magnetic field ${\bf h}^{\rm pulse}(t)$ instead of ${\bf h}^{\rm AC}(t)$ in Eq.~(\ref{eq:def_magneticfield}). 
This enables us to obtain the whole spectrum at once, instead of the study of steady states for each $\omega$. 
We set 
\begin{eqnarray}
&&{\bf h}^{\rm pulse}(t)=
\left\{
\begin{array}{c}
\Delta h\hat{\bm \mu}\quad(0\leq t < 1),\\
0\quad(1\leq t \leq 6000),
\label{eq:hpulse}
\end{array}
\right. 
\end{eqnarray}
where $\Delta h=0.002$ and $\hat{\boldsymbol \mu}$ denotes the unit vector in the direction of $\mu=x,y,z$.
By Fourier transforming the responses, the spectra of $\chi_{E,\mu}(\omega)$ and $\chi_{S,\nu\mu}(\omega)$ are obtained by Eqs.~(\ref{eq:chieef}) and (\ref{eq:chimag}) with the substitution of $h_{\mu}^{\rm AC}(\omega)$ by $h_{\mu}^{\rm pulse}(\omega)$. 
Meanwhile, for the calculations of the spatiotemporal profiles in Secs.~\ref{sec:PBC.3} and \ref{sec:OBC.3}, we apply an AC magnetic field with a frequency $\omega$ in the direction of $\hat{\boldsymbol \mu}$ given by 
\begin{eqnarray}
{\bf h}^{\rm AC}(t)=\Delta h\left(1- e^{-t/t_0}\right)\sin(\omega t)\hat{\boldsymbol \mu} \quad
(0 \leq t \leq 10000), 
\label{eq:hAC}
\end{eqnarray}
where $t_0=50$ is introduced to suppress the initial disturbance by switching on the field.

%%%===========================================================================

%%%===========================================================================
%%%===========================================================================
%%% sec:PBC
%%%===========================================================================
%%%===========================================================================

\section{Result: bulk properties \label{sec:PBC}}

In this section, we show the results of the bulk EEF obtained from the calculations for the system with the PBC.
In Sec.~\ref{sec:PBC.1}, we present the maximum values of $|\chi_{E,\mu}(\omega)|$ on the plane of $h_x^{\rm stat}$ and $h_z^{\rm stat}$, and the typical frequency spectra. 
In Sec.~\ref{sec:PBC.2}, we discuss the relation between the EEF and the magnetic excitations by comparing $\chi_{E,\mu}(\omega)$ and $\chi_{S,\nu\mu}(\omega)$. 
In Sec.~\ref{sec:PBC.3}, we show the spatiotemporal profiles of the spin textures and the EEF 
of the resonance modes.

%%%===========================================================================
%%%sec:PBC.1

\subsection{Complex admittance of the EEF \label{sec:PBC.1}}

%%%---------------------------------------------------------------------------
%%% fig:pbc_chimax

\begin{figure*}[tb]
\centering
\includegraphics[width=1.0\textwidth]{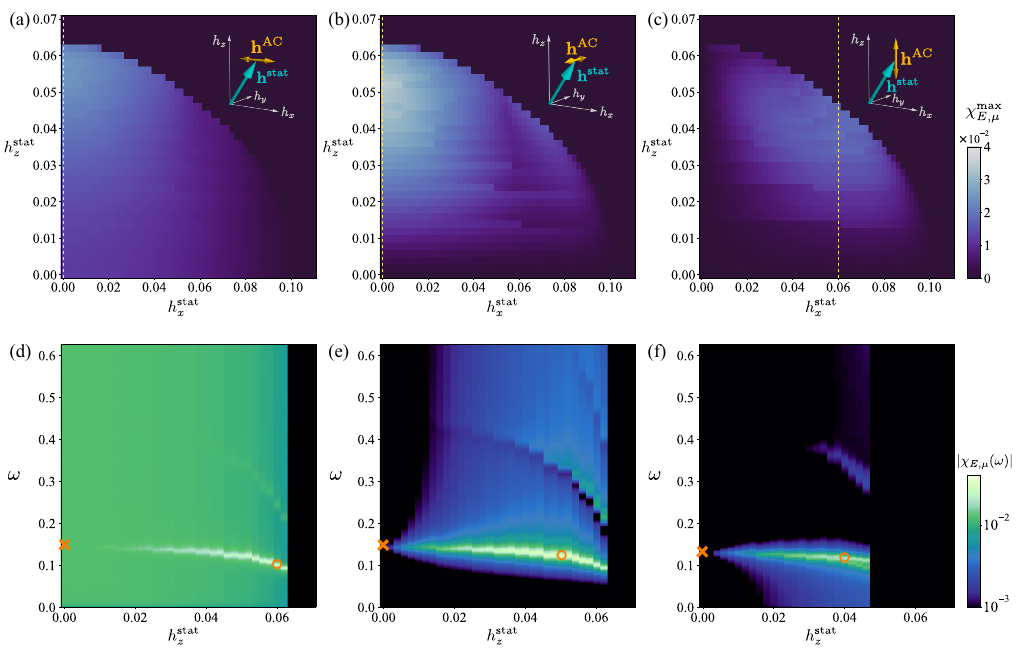}
\caption{
\label{fig:pbc_chimax}
Maximum values of $|\chi_{E,\mu}(\omega)|$, $\chimax{\mu}$, calculated for the model in Eq.~(\ref{eq:ham}) with the PBC for 
(a) ${\bf h}^{\rm AC}\parallel\hat{\bf x}$ ($\mu=x$), (b) ${\bf h}^{\rm AC}\parallel\hat{\bf y}$ ($\mu=y$), and (c) ${\bf h}^{\rm AC}\parallel\hat{\bf z}$ ($\mu=z$), as represented in the insets. 
The frequency dependence of $|\chi_{E,\mu}(\omega)|$ along the yellow dashed lines in (a), (b), and (c) are shown in (d), (e), and (f), respectively: (d) ${\bf h}^{\rm AC}\parallel\hat{\bf x}$ with $h_x^{\rm stat}=0$, (e) ${\bf h}^{\rm AC}\parallel\hat{\bf y}$ with $h_x^{\rm stat}=0$, and (f) ${\bf h}^{\rm AC}\parallel\hat{\bf z}$ with $h_x^{\rm stat}=0.06$. 
The orange circles and crosses in (d), (e), and (f) indicate the parameters for which the real-space behaviors are 
presented in Fig.~\ref{fig:pbc_real}.
}
\end{figure*}
%%%---------------------------------------------------------------------------

We show the results of the complex admittance $\chi_{E,\mu}(\omega)$ in Eq.~\eqref{eq:chieef} for three different directions of the AC magnetic field: $\hacv\parallel \hatv{x}$ ($\mu=x$), $\hacv\parallel \hatv{y}$ ($\mu=y$), and $\hacv\parallel \hatv{z}$ ($\mu=z$). 
Let us first discuss the results for $\hacv\parallel \hatv{x}$. 
In Fig.~\ref{fig:pbc_chimax}(a), we show the maximum values of $|\chi_{E,x}(\omega)|$, denoted by $\chimax{x}$, on the plane of $\hst_x$ and $\hst_z$. 
We find that $\chimax{x}$ has a nonzero value in a region of $0 \leq h_x^{\rm stat} \lesssim 0.103$ and $0 \leq h_z^{\rm stat} \lesssim 0.063$, where the system in the static magnetic field shows a noncollinear spin texture, i.e., the CSL for $h_x^{\rm stat}=0$, conical for $h_z^{\rm stat}=0$, and their mixture otherwise. 
Within this region, $\chimax{x}$ increases with increasing $\hst_z$, but decreases 
with increasing $\hst_x$
; the EEF is maximally generated at $(\hst_x, \hst_z)\simeq(0, 0.06)$, 
which is close to the phase transition from the CSL to the FFM. 
Meanwhile, outside this region, the spins are fully polarized by the magnetic field and $\chi_{E, x}(\omega)$ vanishes, indicating that the FFM phase does not generate the EEF. 

We plot the spectrum of $|\chi_{E,x}(\omega)|$ in Fig.~\ref{fig:pbc_chimax}(d) while changing $\hst_z$ 
at $\hst_x=0$, along the yellow dashed line in Fig.~\ref{fig:pbc_chimax}(a). 
$|\chi_{E,x}(\omega)|$ shows several peaks, on top of an almost $\omega$-independent contribution 
(the origin will be discussed in Sec.~\ref{sec:PBC.2}). 
The peak frequencies decrease but their intensities increase as $\hst_z$ approaches the critical value for the FFM transition, $h_z^{\rm stat, c} \simeq 0.063$. 
$|\chi_{E,x}(\omega)|$ takes the maximum value at the lowest-frequency peak at $\omega \simeq 0.1$, 
which we denote $\omega^{\rm bulk}_1$.
We find the second peak above $\omega^{\rm bulk}_2 \gtrsim 0.2$, but the higher ones are difficult to see in the contour plot [see the $\omega$ profile in Fig.~\ref{fig:pbc_spectra}(a)]. 
These peak structures are related to magnetic resonance, as will be discussed in Sec.~\ref{sec:PBC.2}.

Next, we discuss $\chimax{y}$ for $\hacv\parallel\hatv{y}$ shown in Fig.~\ref{fig:pbc_chimax}(b). 
The EEF is again maximally generated in the CSL phase for $h_x^{\rm stat}=0$, but at a slightly lower $h_z^{\rm stat}\simeq 0.05$ below the critical value for the FFM transition. 
The intensity is stronger than that in Fig.~\ref{fig:pbc_chimax}(a). 
We find another weaker peak at $(\hst_x, \hst_z) \simeq (0.07, 0.04)$, 
which is a remnant of the peak for the case of $\hacv\parallel\hatv{z}$ discussed below for Fig.~\ref{fig:pbc_chimax}(c). 
We note that $\chimax{y}$ is strongly reduced in the conical state at $\hst_z=0$ in contrast to the case of $\hacv\parallel \hatv{x}$ in Fig.~\ref{fig:pbc_chimax}(a). 
This is due to twofold rotational symmetry about the $x$ axis accompanied by time translation by a half period of ${\bf h}^{\rm AC}$, $\frac{\pi}{\omega}$~\cite{3a_note}.
The spectrum of $|\chi_{E,y}(\omega)|$ for $\hst_x=0$ [the yellow dashed line in Fig.~\ref{fig:pbc_chimax}(b)] is shown in Fig.~\ref{fig:pbc_chimax}(e). 
We can identify three sharp peak structures within this $\omega$ range, rather more easily than in Fig.~\ref{fig:pbc_chimax}(d). 
We note that, in contrast to the case of Fig.~\ref{fig:pbc_chimax}(d), there is no $\omega$-independent contribution, and that the peak intensities are maximized at $\hst_z \simeq 0.05$, and decreased for $\hst_z \gtrsim 0.05$. 

Lastly, we show the results for $\hacv\parallel\hatv{z}$ in Figs.~\ref{fig:pbc_chimax}(c) and \ref{fig:pbc_chimax}(f). 
In this case, the EEF is maximally generated at $(\hst_x, \hst_z) \simeq (0.06, 0.045)$, which is close to the subdominant maximum in Fig.~\ref{fig:pbc_chimax}(b). 
Similar to the case of Fig.~\ref{fig:pbc_chimax}(b), $\chimax{z}$ is strongly suppressed for $\hst_z=0$ by symmetry~\cite{3a_note}. 
Moreover, in contrast to the cases of Figs.~\ref{fig:pbc_chimax}(a) and \ref{fig:pbc_chimax}(b), $\chimax{z}$ is zero for $\hst_x=0$ due to twofold rotational symmetry about the $z$ axis. 
Under these symmetric constraints, the peak of $\hacv\parallel\hatv{z}$ appears in the intermediate region near the phase boundary to the FFM state. 
Figure~\ref{fig:pbc_chimax}(f) shows $|\chi_{E,z}(\omega)|$ for $\hst_x=0.06$, along the yellow dashed line through the peak of $\chimax{z}$ in Fig.~\ref{fig:pbc_chimax}(c). 
We find several peaks without an $\omega$-independent background, as in the case of Fig.~\ref{fig:pbc_chimax}(e). 
We note that each peak splits into two, as most clearly seen at the lowest-frequency peak $\omega^{\rm bulk}_1$. 
This is understood from the magnon dispersion for the noncoplanar spin state, but the details will be discussed elsewhere.

%%%===========================================================================

%%%===========================================================================
%%% sec: PBC.2

\subsection{Comparison with magnetic excitations \label{sec:PBC.2}}

%%%---------------------------------------------------------------------------
%%% fig:pbc_spectra
\begin{figure}[tb]
\centering
\includegraphics[width=1.0\columnwidth]{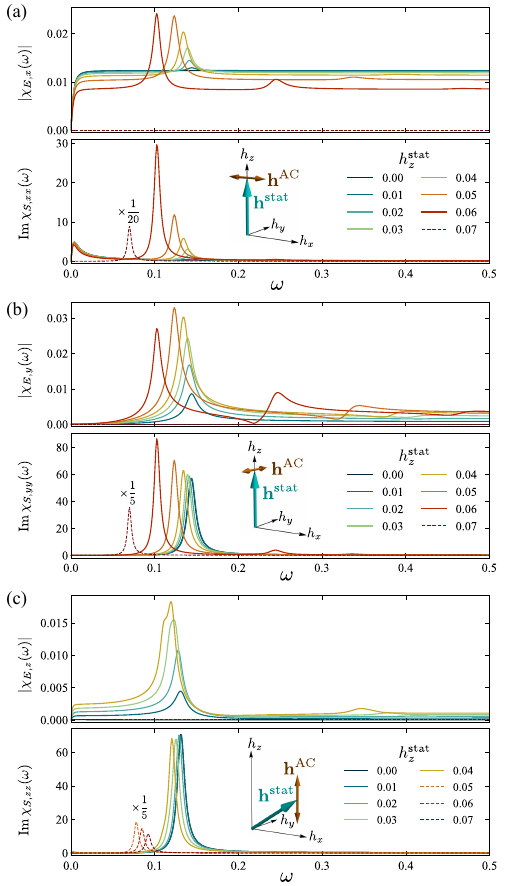}
\caption{
\label{fig:pbc_spectra}
Comparison between $|\chi_{E,\mu}(\omega)|$ and the imaginary part of the dynamical spin susceptibility, $\ImchiS{\mu\mu}$, for several values of $h_z^{\rm stat}$: (a) ${\bf h}^{\rm AC}\parallel\hat{\bf x}$ with $h_x^{\rm stat}=0$, 
(b) ${\bf h}^{\rm AC}\parallel\hat{\bf y}$ with $h_x^{\rm stat}=0$, and (c) ${\bf h}^{\rm AC}\parallel\hat{\bf z}$ with $h_x^{\rm stat}=0.06$ for the model in Eq.~(\ref{eq:ham}) with the PBC. 
The data for the FFM phases are denoted by the dotted lines and scaled for better visibility. 
}
\end{figure}
%%%---------------------------------------------------------------------------

In this section, we discuss the results of $|\chi_{E,\mu}(\omega)|$ in comparison with the magnetic excitation spectrum. 
Figure~\ref{fig:pbc_spectra} displays the comparisons between $|\chi_{E,\mu}(\omega)|$ and the imaginary part of the dynamical spin susceptibility, $\ImchiS{\mu\mu}$, calculated by Eq.~\eqref{eq:chimag} for different field configurations: 
$\hacv\parallel\hatv{x}$ ($\mu=x$) and $\hst_x=0$ in Fig.~\ref{fig:pbc_spectra}(a), 
$\hacv\parallel\hatv{y}$ ($\mu=y$) and $\hst_x=0$ in Fig.~\ref{fig:pbc_spectra}(b), and 
$\hacv\parallel\hatv{z}$ ($\mu=z$) and $\hst_x=0.06$ in Fig.~\ref{fig:pbc_spectra}(c) 
[on the yellow dashed lines in Figs.~\ref{fig:pbc_chimax}(a), \ref{fig:pbc_chimax}(b), and \ref{fig:pbc_chimax}(c), respectively]. 
In each figure, the different colors represent the different values of $\hst_z$. 
The solid lines denote the results in the noncollinear spin phase where $|\chi_{E,\mu}(\omega)|$ is nonzero in Figs.~\ref{fig:pbc_chimax}(a)-\ref{fig:pbc_chimax}(c), while the dashed lines denote the results in the outside, i.e., the FFM phase where $|\chi_{E,\mu}(\omega)|=0$. 

Let us first discuss the results for $\hacv\parallel\hatv{x}$ and $\hst_x=0$. 
As shown in Fig.~\ref{fig:pbc_spectra}(a), 
$|\chi_{E,x}(\omega)|$ and $\ImchiS{xx}$ have peaks at the same frequencies, except for $\hst_z = 0.07$ where the system is in the FFM phase and $|\chi_{E,x}(\omega)|$ vanishes. 
This indicates that the EEF is amplified at the magnetic resonance frequencies in the noncollinear spin phase. 
The resonance frequencies decrease as $\hst_z$ increases, which is consistent with the previous studies for the magnetic resonance in the CSL phase~\cite{Kishine2009,Kiselev2013}. 
Interestingly, the higher-$\omega$ peaks are more clearly visible in $|\chi_{E,x}(\omega)|$ 
rather than those in $\ImchiS{xx}$. 
This indicates that the EEF response has an advantage to observe the higher-$\omega$ excitations than the magnetic one. 
This is presumably due to an additional $\omega$-linear factor from $\frac{\partial {\bf S}_l(t)}{\partial t}$ in Eq.~(\ref{eq:eefloc}); see also Eq.~\eqref{eq:app_chiEres_h=0} in Appendix~\ref{appendix:chiE_resonance}. 

While $|\chi_{E,x}(\omega)|$ accompanies an almost $\omega$-independent contribution (see below), the peak heights measured from it grow from zero as $\hst_z$ increases from zero, in good correspondence with the growth of the peaks in $\ImchiS{xx}$. 
We note that the growth rate of the peaks in $|\chi_{E,x}(\omega)|$ decreases as 
$\hst_z$ approaches the critical field $h_z^{\rm stat, c}$, in contrast to the rapid increase of the growth rate in $\ImchiS{xx}$. 
This feature is qualitatively understood from the decreases of $\omega_1^{\rm bulk}$ and the vector spin chirality that bridges $\chi_{E,x}(\omega)$ and $\chi_{S,xx}$. 
Here, the vector chirality in the ground state is defined as 
\begin{eqnarray}
\bar{\bf C}^{\rm vc} 
= \frac{1}{L}\sum_{l=0}^{L-1}{\bf S}_l(0)\times{\bf S}_{l+1}(0). 
\label{eq:vector_chirality}
\end{eqnarray}
This quantity measures the overall spin noncollinearity, and hence is rapidly reduced as $\hst_z$ approaches $h_z^{\rm stat, c}$ where the spin texture acquires a strong solitonic feature. 
This and the decrease of $\omega_1^{\rm bulk}$ suppress the growth rate of the peaks in $|\chi_{E,x}(\omega)|$;  see Eq.~\eqref{eq:app_chiEres_h=0} in Appendix~\ref{appendix:chiE_resonance}.

The $\omega$-independent contributions in $|\chi_{E,x}(\omega)|$ 
is explained by the consideration of the large-$\omega$ behavior as follows. 
For larger $\omega$, we can derive the relation 
\begin{eqnarray}
\chi_{E,\mu}(\omega) \simeq -\frac{\alpha}{1+\alpha^2}
\bar{\bf C}^{\rm vc}
\cdot
\hat{\boldsymbol{\mu}}. 
\label{eq:chiinf}
\end{eqnarray}
See Appendix~\ref{appendix:chiE_limit} for the derivation. 
Thus, the large-$\omega$ behavior is independent of $\omega$, and it is given by the overall spin noncollinearity $\bar{\bf C}^{\rm vc}\cdot\hat{\bf x}$. 
Since $\bar{\bf C}^{\rm vc}\cdot\hat{\bf x}$ decreases with increasing $\hst_z$, the $\omega$-independent component of $|\chi_{E,x}(\omega)|$ is reduced, as shown in the upper panel of Fig.~\ref{fig:pbc_spectra}(a). 
We note that this contribution corresponds to the collective dynamics of spin textures called coherent sliding dynamics~\cite{Kishine2012}. 
We will return to this point in Sec.~\ref{sec:PBC.3}.

Next, we discuss the results for $\hacv\parallel\hatv{y}$ and $\hst_x=0$.
In this case also, $|\chi_{E,y}(\omega)|$ shows peaks corresponding to the magnetic resonances in $\ImchiS{yy}$, 
and the higher-$\omega$ modes are more clearly visible in $|\chi_{E,y}(\omega)|$ than $\ImchiS{yy}$, as shown in Fig.~\ref{fig:pbc_spectra}(b). 
In contrast to the case with $\hacv\parallel\hatv{x}$, 
however, the peak heights in $|\chi_{E,y}(\omega)|$ increase for $0 \leq \hst_z \lesssim 0.055$ but decrease  
for $\hst_z \gtrsim 0.055$, 
while $\ImchiS{yy}$ shows resonance peaks even for $\hst_z=0$ and grows monotonically with $\hst_z$. 
This nonmonotonic behavior of $|\chi_{E,y}(\omega)|$ is already observed in Fig.~\ref{fig:pbc_chimax}(b).  
The decrease of $|\chi_{E,y}(\omega)|$ can be ascribed to the slow increase of $\ImchiS{yy}$ in the lower panel of Fig.~\ref{fig:pbc_spectra}(b), in addition to the decreases of $\bar{C}^{\rm vc}_x$ and $\omega_1^{\rm bulk}$ discussed above. 
Furthermore, there is no $\omega$-independent contributions, since $\bar{C}^{\rm vc}_y=0$ in Eq.~\eqref{eq:vector_chirality}, leading to $ \chi_{E,y}(\omega) \simeq 0$ in Eq.~\eqref{eq:chiinf} for large $\omega$~\footnote{
The long tails of $|\chi_{E,y}(\omega)|$ in the large-$\omega$ region are due to the discretization in the numerical computation of the spatial derivative in Eq.~\eqref{eq:eefloc}.
}.

In Fig.~\ref{fig:pbc_spectra}(c), we show the results for $\hacv\parallel\hatv{z}$ and $\hst_x=0.06$. 
Similar to the above two cases, the higher-$\omega$ peaks are more clearly visible for $|\chi_{E,z}(\omega)|$ than $\ImchiS{zz}$. 
The lowest-energy peak in $|\chi_{E,z}(\omega)|$, which shows a shoulder-like feature reflecting the splitting mentioned in Sec.~\ref{sec:PBC.1}, grows with $\hst_z$, as in the case of $\hacv\parallel\hatv{x}$. 
This is understood from the increase of $\bar{C}_z^{\rm vc}$  
with $\hst_z$; see again Eq.~\eqref{eq:app_chiEres_h=0} in Appendix~\ref{appendix:chiE_resonance}. 
In contrast, the peak heights in $\ImchiS{zz}$ are almost unchanged for $\hst_z$ since the solitonic feature in the spin texture is not well developed in this range of $\hst_z$.

%%%===========================================================================

%%%===========================================================================
%%% sec: PBC.3

\subsection{Real-space behavior \label{sec:PBC.3}}

%%%---------------------------------------------------------------------------
%%% fig:pbc_real
\begin{figure*}[tb]
\centering
\includegraphics[width=1.0\textwidth]{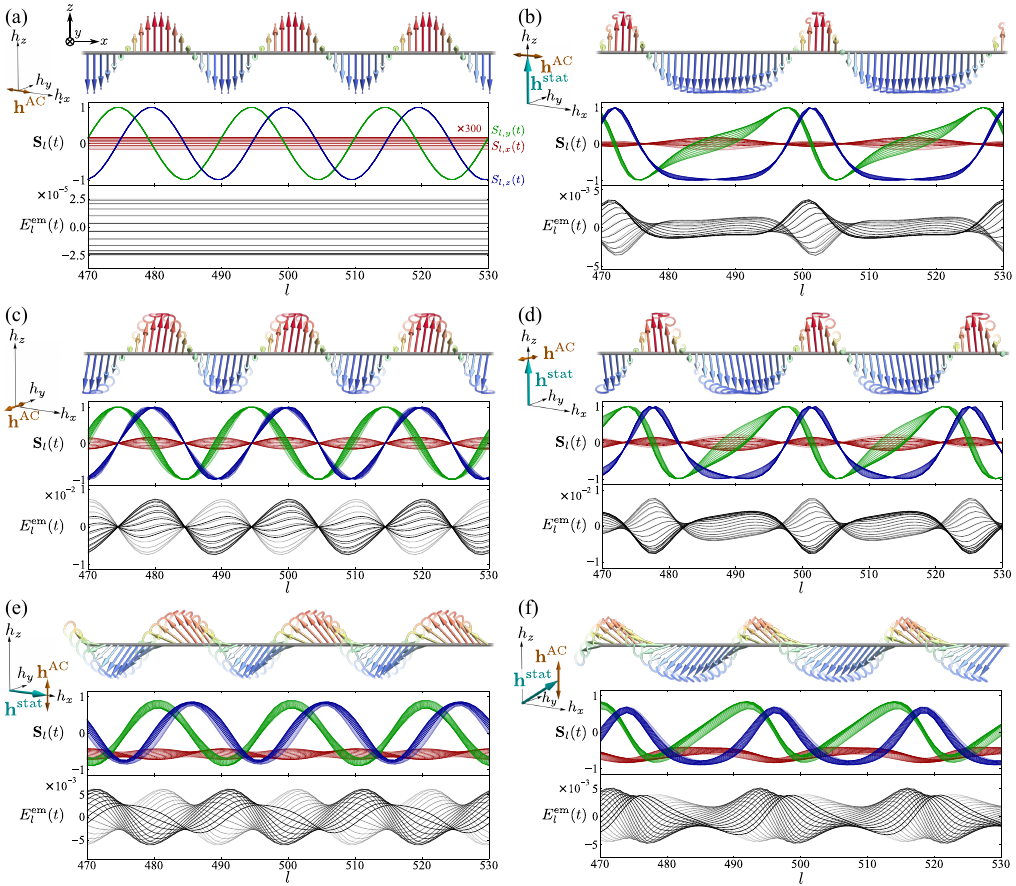}
\caption{
\label{fig:pbc_real}
Real-space spin configurations (top and middle) and real-space distributions of the EEF (bottom) in the steady states under the AC magnetic field ${\bf h}^{\rm AC}$. 
(a) and (b), (c) and (d), and (e) and (f) are for ${\bf h}^{\rm AC}\parallel\hat{\bf x}$, ${\bf h}^{\rm AC}\parallel\hat{\bf y}$, and ${\bf h}^{\rm AC}\parallel\hat{\bf z}$, respectively, at 
(a) $(h_x^{\rm stat}, h_z^{\rm stat})=(0,0)$ and $\omega=0.143$, 
(b) $(h_x^{\rm stat}, h_z^{\rm stat})=(0,0.06)$ and $\omega=0.103$, 
(c) $(h_x^{\rm stat}, h_z^{\rm stat})=(0,0)$ and $\omega=0.143$, 
(d) $(h_x^{\rm stat}, h_z^{\rm stat})=(0,0.05)$ and $\omega=0.124$, 
(e) $(h_x^{\rm stat}, h_z^{\rm stat})=(0.06,0)$ and $\omega=0.131$, and 
(f) $(h_x^{\rm stat}, h_z^{\rm stat})=(0.06,0.04)$ and $\omega=0.123$. 
All the frequencies are set at the values for the lowest-$\omega$ resonance mode; 
the values are denoted by the orange crosses and circles in Figs.~\ref{fig:pbc_chimax}(d)--\ref{fig:pbc_chimax}(f). 
The results are shown for $60$ sites at the center of the $1000$-site system with the PBC.
In each figure, the top panel displays the spin configurations by the arrows with the lines representing the trajectories of the arrowheads during a single period of ${\bf h}^{\rm AC}$.
Meanwhile, the middle panel displays the spin components $S_{l, x}(t)$, $S_{l,y}(t)$, and $S_{l,z}(t)$ by the red, green, and blue lines, respectively, and the bottom panel displays the EEF at each site, $E^{\rm em}_l(t)$, during the corresponding period of ${\bf h}^{\rm AC}$. 
The color and gray-scale intensities in the middle and bottom panels, respectively, increase with time evolution in the single period of ${\bf h}^{\rm AC}$. 
In the middle panel of (a), $S_{l,x}(t)$ is multiplied by a factor of 300 for better visibility. 
}
\end{figure*}
%%%---------------------------------------------------------------------------

%%%---------------------------------------------------------------------------
%%% fig:pbc_xt
\begin{figure*}[tb]
\centering
\includegraphics[width=1.0\textwidth]{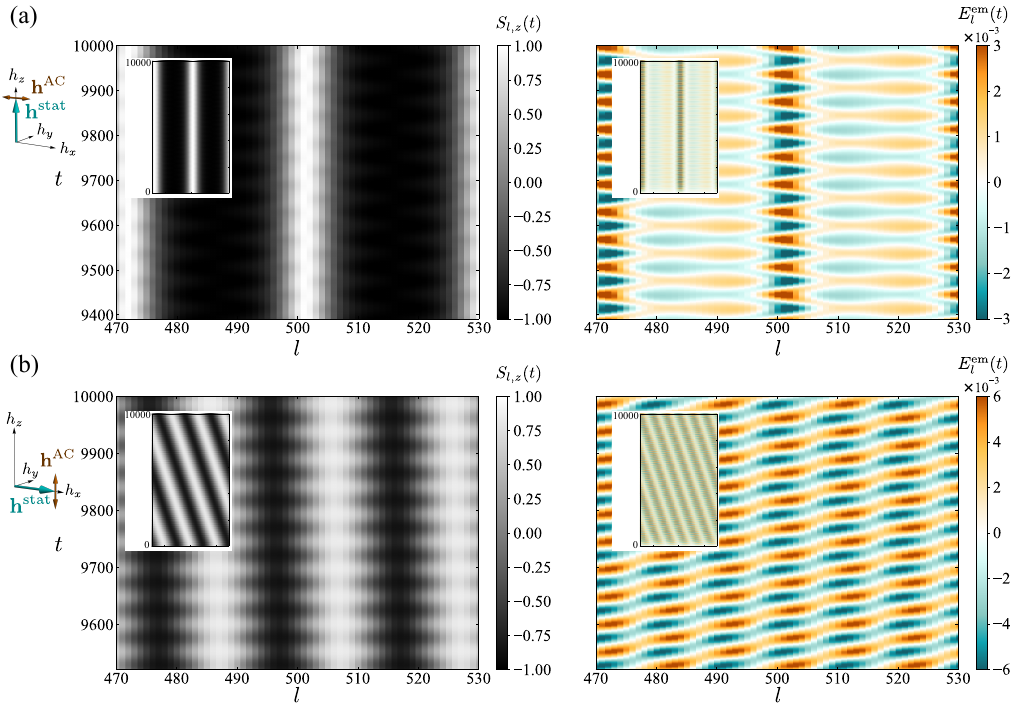}
\caption{
\label{fig:pbc_xt}
Real-space and real-time distribution of the $z$ component of spins $S_{l,z}(t)$ (left) and $E^{\rm em}_l(t)$ (right) 
within 10 periods of ${\bf h}^{\rm AC}$ on the plane of $l$ and $t$: 
(a) and (b) correspond to Figs.~\ref{fig:pbc_real}(b) and \ref{fig:pbc_real}(e), respectively.
The insets show the results for a longer time window.
}
\end{figure*}
%%%---------------------------------------------------------------------------

In this section, we present the spatiotemporal profiles of spins ${\bf S}_l(t)$ and EEF $E_l^{\rm em}(t)$ in Eq.~\eqref{eq:eefloc}. 
Figure~\ref{fig:pbc_real} displays the results for the lowest-$\omega$ resonance mode: 
The data are calculated for the parameters indicated by the circles and crosses in Figs.~\ref{fig:pbc_chimax}(d)--\ref{fig:pbc_chimax}(f). 
In each figure of Fig.~\ref{fig:pbc_real}, the top panel shows the spin precession motions, the middle panel shows the time evolution of each spin component, and the bottom panel shows the time evolution of $E_l^{\rm em}(t)$ during a single period of ${\bf h}^{\rm AC}$ for $60$ sites at the center of the $1000$-site system with the PBC.  

First, we discuss the results for $\hacv\parallel\hatv{x}$ and $\hst_x=0$, corresponding to Fig.~\ref{fig:pbc_chimax}(d). 
Figure~\ref{fig:pbc_real}(a) is for the helical state at $(\hst_x, \hst_z)=(0,0)$ and $\omega=0.143$, indicated by the cross in Fig.~\ref{fig:pbc_chimax}(d). 
In this case, the spin dynamics is strongly suppressed, leaving very weak oscillations, as visible in the enlarged plot of $S_{l,x}(t)$ in the middle panel of Fig.~\ref{fig:pbc_real}(a). 
We note that such oscillations are induced not only at the resonance frequency but also for general $\omega$, and they are called the coherent sliding dynamics~\cite{Kishine2012}. 
In this situation, we obtain a small but spatially uniform $E_l^{\rm em}(t)$, as shown in the bottom panel of Fig.~\ref{fig:pbc_real}(a). 
The spatial average in Eq.~\eqref{eq:eef}, however, can be comparably large to those for the other resonance modes, which leads to the $\omega$-independent contribution in Fig.~\ref{fig:pbc_spectra}(a). 

Figure~\ref{fig:pbc_real}(b) is also for $\hacv\parallel\hatv{x}$, but at $(\hst_x, \hst_z)=(0,0.06)$ and $\omega=0.103$, indicated by the cirlcle in Fig.~\ref{fig:pbc_chimax}(d). 
In contrast to the above helical case, the spin dynamics in this CSL state is resonantly activated, especially in the regions between the solitons with $S_{l, z}(t) \simeq +1$, as shown in the top and middle panels of Fig.~\ref{fig:pbc_real}(b). 
In contrast, the EEF is largely generated around the solitons, as shown in the bottom panel. 
This trend is understood from the fact that the vector spin chirality is large near the solitons. 
The spatially averaged EEF takes a large value, leading to the resonance peak of $|\chi_{E,x}(\omega)|$ in Figs.~\ref{fig:pbc_chimax}(d) and \ref{fig:pbc_spectra}(a). 
We note that in this resonance with $\hst_x=0$, the spin texture is just oscillating around the original position in real space, and the EEF behaves like a standing wave, as shown in Fig.~\ref{fig:pbc_xt}(a).  
This is due to twofold rotational symmetry about the $z$ axis with time translation by $\frac{\pi}{\omega}$. 

Next, we discuss the results for $\hacv\parallel\hatv{y}$ and $\hst_x=0$, corresponding to Fig.~\ref{fig:pbc_chimax}(e).  
Figure~\ref{fig:pbc_real}(c) is for the helical state at $\hst_z=0$ and $\omega=0.143$, indicated by the cross in Fig.~\ref{fig:pbc_chimax}(e). 
In this case, the spins pointing in the $\pm y$ directions do not precess and constitute the nodes of oscillating ${\bf S}_l(t)$, and those pointing in the $\pm z$ directions are the antinodes, where the spin precessions become maximum. 
The EEF vanishes at the nodes, whereas it is maximally generated at the antinodes reflecting the large spin precessions. 
Note, however, that the average of the EEF is strongly suppressed due to twofold rotational symmetry about the $x$ axis accompanied by $\frac{\pi}{\omega}$ time translation of $\hacv$~\cite{3a_note}. 
Meanwhile, Fig.~\ref{fig:pbc_real}(d) is for a CSL state at $\hst_z=0.05$ with $\omega=0.124$ corresponding to the circle in Fig.~\ref{fig:pbc_chimax}(e). 
In this case, in the regions between (near) the solitons, the amplitude of the spin precessions is enhanced (suppressed), while $E_{l}^{\rm em}(t)$ is suppressed (slightly enhanced). 
The spatially averaged $E_{l}^{\rm em}(t)$ leads to the sharp resonance peak in Figs.~\ref{fig:pbc_chimax}(e) and \ref{fig:pbc_spectra}(b). 
In these cases with  $\hst_x=0$ also, the spin texture and the EEF behave like standing waves similar to those in Fig.~\ref{fig:pbc_xt}(a) because of the symmetry. 

Finally, we discuss the results for $\hacv\parallel\hatv{z}$; here we take $\hst_x=0.06$ since  $\bar{E}^{\rm em}(t)$ vanishes for $\hst_x=0$, as explained in Sec.~\ref{sec:PBC.1}. 
Figure~\ref{fig:pbc_real}(e) is for the conical state at $\hst_z=0$ and $\omega=0.131$, indicated by the cross in Fig.~\ref{fig:pbc_chimax}(f). 
In this case, the spin precessions become larger (smaller) for $S_{l,z}(t) \simeq 0$ [$S_{l, y}(t) \simeq 0$], since the AC magnetic field is applied along the $z$ direction. 
In contrast to the above cases with  $\hst_x=0$, the spin texture is driven from right to left, as shown in the left panel of Fig.~\ref{fig:pbc_xt}(b); the velocity is estimated as $v \simeq -0.0035$. 
This drift motion is known as the Archimedean screw dynamics~\cite{Nina2021}. 
Correspondingly, $E^{\rm em}_l(t)$ is largely generated and no longer behaves like a standing wave. 
Reflecting the drift motion, the EEF is also driven from right to left with the same velocity, as shown in the inset of Fig.~\ref{fig:pbc_xt}(b).
In addition, we find that the EEF propagates from left to right with much faster velocity, 
as shown in the enlarged plot in the main panel of Fig.~\ref{fig:pbc_xt}(b) as well as the bottom panel of Fig.~\ref{fig:pbc_real}(e). 
In this faster mode, the EEF propagates from one soliton to the next one during one cycle of time; 
hence, the velocity is estimated as $v' \simeq \frac{\omega}{Q_0} \simeq 0.42$, where $Q_0$ is the ordering wave number. 
This is more than 100 times faster than $v$. 
Such a fast motion is hardly seen in $S_{l,y}(t)$ and $S_{l, z}(t)$, but discernible in $S_{l,x}(t)$ 
in the middle panel of Fig.~\ref{fig:pbc_real}(e).  
Since the motion in $S_{l,x}(t)$ corresponds to spin precession due to magnon excitations in the conical state, the fast propagation of the EEF to the field direction is associated with the magnon propagation. 
Thus, the parallel component of the static magnetic field, $\hst_x$, activates two types of propagating motions of the EEF: the fast motion to the field direction due to the magnon propagation and the slow drift to the opposite direction due to the Archimedean screw dynamics of the spin texture.

Although the spatially averaged EEF $\bar{E}^{\rm em}(t)$ vanishes in the case with $\hst_z=0$, it becomes nonzero for nonzero $\hst_z$, as a solitonic feature is induced in the spin texture. 
Such results are shown in Fig.~\ref{fig:pbc_real}(f) for $\hst_z=0.04$ and $\omega=0.123$ corresponding to the circle in Fig.~\ref{fig:pbc_chimax}(f). 
In this case also, the spin texture and the EEF is driven to left associated with the Archimedean screw dynamics, while the EEF also propagates to right with much faster velocity associated with the magnon propagation, similar to Fig.~\ref{fig:pbc_xt}(b).

%%%===========================================================================

%%%===========================================================================
%%%===========================================================================
%%% sec:OBC
%%%===========================================================================
%%%===========================================================================

\section{Result: Edge contributions \label{sec:OBC}}

%%%---------------------------------------------------------------------------
%%% fig:obc_chimax
\begin{figure*}[tb]
\centering
\includegraphics[width=1.0\textwidth]{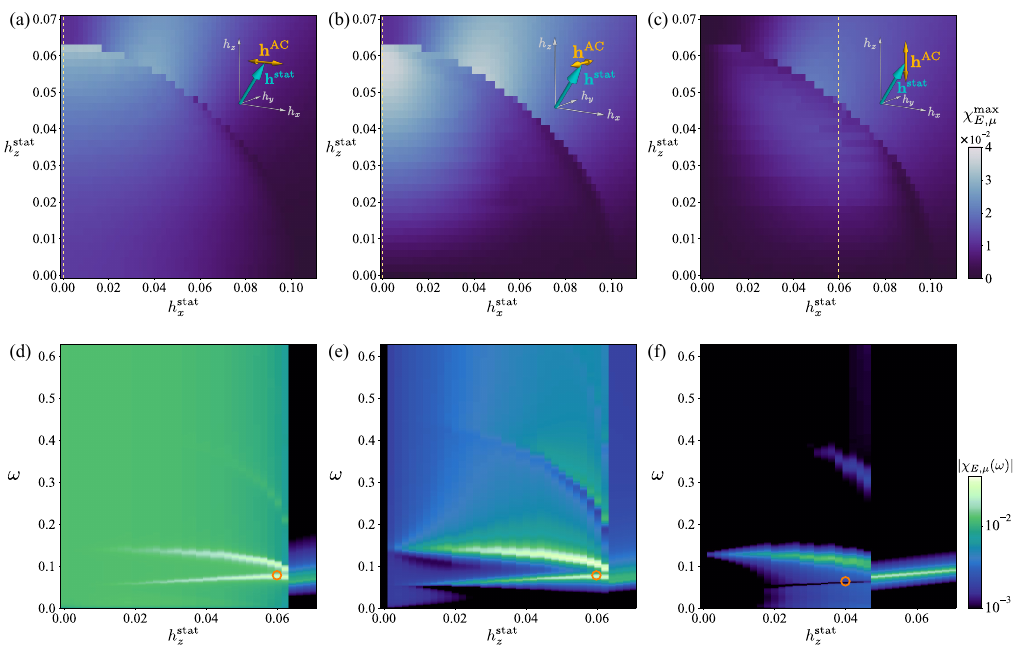}
\caption{
\label{fig:obc_chimax}
Maximum value of $|\chi_{E,\mu}(\omega)|$, $\chimax{\mu}$, calculated for the model in Eq.~\eqref{eq:ham} with the OBC for 
(a) ${\bf h}^{\rm AC}\parallel\hat{\bf x}$ ($\mu=x$), 
(b) ${\bf h}^{\rm AC}\parallel\hat{\bf y}$ ($\mu=y$), and 
(c) ${\bf h}^{\rm AC}\parallel\hat{\bf z}$ ($\mu=z$). 
The system size is $L=1000$.
The frequency dependence of $|\chi_{E,\mu}(\omega)|$ along the yellow dashed lines in (a), (b), and (c) are shown in (d), (e), and (f), respectively: 
(d) $\hacv\parallel\hatv{x}$ with $\hst_x=0$, 
(e) $\hacv\parallel\hatv{y}$ with $\hst_x=0$, and 
(f) $\hacv\parallel\hatv{z}$ with $\hst_x=0.06$. 
The orange circles in (d), (e), and (f) indicate the parameters for which the real-space behaviors are presented in Fig.~\ref{fig:obc_real}.
}
\end{figure*}
%%%---------------------------------------------------------------------------

Thus far, we have shown the results for the bulk contributions in the system with the PBC. 
In this section, we turn our attention to the contributions from the edges of the system by employing the OBC. 
In Sec.~\ref{sec:OBC.1}, we first present the results of  $|\chi_{E,\mu}(\omega)|$ to show the edge contributions to the EEF through additional resonance modes. 
In Sec.~\ref{sec:OBC.2}, we discuss the difference between the bulk resonance modes and the additional modes based on the spectra of $|\chi_{E,\mu}(\omega)|$ and $\ImchiS{\mu\mu}$, including the system size dependence. 
In Sec.~\ref{sec:OBC.3}, we show the spatiotemporal profiles of the spin textures and the EEF for the additional resonance modes to explicitly show that these modes are localized at the edges of the system.

%%%===========================================================================
%%% sec: OBC.1
\subsection{Complex admittance of the EEF \label{sec:OBC.1}}

As in Sec.~\ref{sec:PBC.1}, we show the results of the complex admittance $\chi_{E,\mu}(\omega)$ for the system with the OBC. 
First, we discuss the results for $\hacv\parallel\hatv{x}$ shown in Figs.~\ref{fig:obc_chimax}(a) and \ref{fig:obc_chimax}(d). 
Similar to the PBC case, in the noncollinear spin phase, $\chimax{x}$ increases as $\hst_x$ decreases and $\hst_z$ increases, showing the maximum at $(\hst_x, \hst_z) \simeq (0, 0.06)$. 
Notably, however, $\chimax{x}$ becomes nonzero also in the FFM phase outside of this region, being comparably large near the phase boundary at $(\hst_x, \hst_z) \simeq (0.04, 0.055)$. 
This contribution originates purely from the edges of the system with the OBC since it was absent for the PBC case in Fig.~\ref{fig:pbc_chimax}(a). 
In addition, we find an additional resonance mode in the noncollinear spin phase, as shown in the spectrum of $|\chi_{E,x}(\omega)|$ in Fig.~\ref{fig:obc_chimax}(d) along the yellow dashed line in Fig.~\ref{fig:obc_chimax}(a); besides several peaks on top of an almost $\omega$-independent contribution already present for the PBC case in Fig.~\ref{fig:pbc_chimax}(d), $|\chi_{E,x}(\omega)|$ shows a sharp peak at a lower $\omega$. 
The resonance frequency of this additional mode, which we denote $\omega^{\rm edge}$, increases as $\hst_z$ increases, in contrast to the bulk resonance frequencies $\omega_n^{\rm bulk}$. 
This is the edge mode reported in the previous study~\cite{Hoshi2020}. 
We will show that the intensity at $\omega^{\rm edge}$ depends on the system size and that the spin dynamics and the EEF associated with this resonance mode are localized near the edges of the system in the following sections.

The situation is similar for the case of $\hacv \parallel \hatv{y}$ shown in Figs.~\ref{fig:obc_chimax}(b) and \ref{fig:obc_chimax}(e); 
we obtain an additional resonance mode at $\omega^{\rm edge}$, in addition to nonzero contributions even in the FFM phase. 
In contrast, for $\hacv\parallel\hatv{z}$ in Figs.~\ref{fig:obc_chimax}(c) and \ref{fig:obc_chimax}(f), instead of the additional resonance peak, we find a sharp dip at $\omega^{\rm edge}$ on top of the long tail of $\omega_1^{\rm bulk}$. 
This is due to the fact that $\chi_{E,z}(\omega)$ from the edges has an opposite sign to that for the bulk. 
We also note that $|\chi_{E,z}(\omega)|$ shows a sharper peak in the FFM phase, compared to the other cases in Figs.~\ref{fig:obc_chimax}(d) and \ref{fig:obc_chimax}(e).

%%%===========================================================================

%%%===========================================================================
%%% sec: OBC.2

\subsection{Comparison with magnetic excitations \label{sec:OBC.2}}

%%%---------------------------------------------------------------------------
%%% fig:obc_spectra
\begin{figure}[tb]
\centering
\includegraphics[width=1.0\columnwidth]{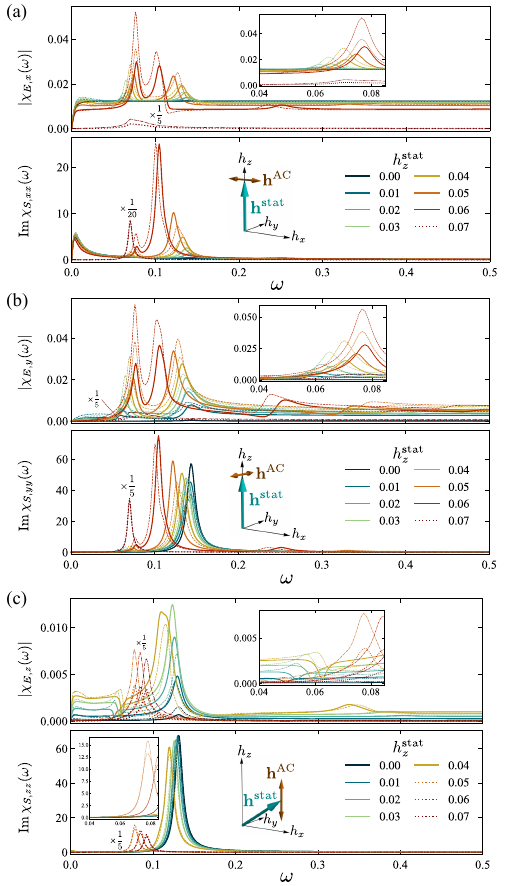}
\caption{
\label{fig:obc_spectra}
Comparison between $|\chi_{E,\mu}(\omega)|$ and the imaginary part of the dynamical spin susceptibility, $\ImchiS{\mu\mu}$, for several values of $h_z^{\rm stat}$: (a) ${\bf h}^{\rm AC}\parallel\hat{\bf x}$ with $h_x^{\rm stat}=0$, 
(b) ${\bf h}^{\rm AC}\parallel\hat{\bf y}$ with $h_x^{\rm stat}=0$, and (c) ${\bf h}^{\rm AC}\parallel\hat{\bf z}$ with $h_x^{\rm stat}=0.06$ for the model in Eq.~(\ref{eq:ham}) with the OBC. 
The solid and dashed (dotted and dashed-dotted) lines denote the data in the noncollinear spin (FFM) phase with $L=1000$ and $L=500$, respectively. 
The data for the FFM phases are scaled for better visibility. 
}
\end{figure}
%%%---------------------------------------------------------------------------

Next, we compare the results of $|\chi_{E,\mu}(\omega)|$ with $\ImchiS{\mu\mu}$ for the system with the OBC, as in Sec.~\ref{sec:PBC.2}. 
Let us first discuss the results for $\hacv\parallel\hatv{x}$ and $\hst_x=0$ shown in Fig.~\ref{fig:obc_spectra}(a); the solid and dashed lines denote the results with $L=1000$ and $L=500$, respectively, in the noncollinear spin phase for $\hst_z \lesssim 0.063$, while the dotted and dashed-dotted lines denote the results with $L=1000$ and $L=500$, respectively, in the FFM phase for $\hst_z \gtrsim 0.063$. 
In the noncollinear spin phase, both $|\chi_{E,x}(\omega)|$ and $\ImchiS{xx}$ exhibit a resonance peak at $\omega^{\rm edge} \simeq 0.06 - 0.08$, 
in addition to the bulk resonance peaks at higher $\omega^{\rm bulk}_n$ and the $\omega$-independent contribution already observed in the PBC case. 
We will show explicitly that this additional contribution originates from the localized modes at the edges in the next section. 
Accordingly, the edge contribution shows conspicuous system size dependences: The peak heights for $L=1000$ become about twice as small as those for $L=500$, while the bulk ones do not change largely. 
As $|\chi_{E,x}(\omega)|$ describes the response averaged over the system, the result indicates that the edge contribution is almost system size independent, while the bulk one is proportional to the system size. 
Notably, the intensity of the edge resonance peak of $|\chi_{E,x}(\omega)|$ increases with $\hst_z$ and can be stronger than the bulk ones, even though that of $\ImchiS{xx}$ is much weaker than the bulk ones, as shown in the lower panel of Fig.~\ref{fig:obc_spectra}(a).  
We also find that $|\chi_{E,x}(\omega)|$ shows a broad peak in the FFM phase for $\hst_z \gtrsim 0.063$, 
in contrast to the PBC case. 
 
Similar results are obtained for $\hacv\parallel\hatv{y}$, as shown in Fig.~\ref{fig:obc_spectra}(b). 
Meanwhile, for $\hacv\parallel\hatv{z}$, $|\chi_{E,z}(\omega)|$ shows a dip at $\omega^{\rm edge}$ 
as shown in Fig.~\ref{fig:obc_spectra}(c), as a result of a cancellation between the long tail of the bulk response at $\omega^{\rm bulk}_1$ and the edge resonance contribution with the opposite sign. 
This is confirmed by the observation that the dip becomes almost twice shallower for $L=1000$ compared to that for $L=500$. 
We note that the edge contribution is much weaker in both $|\chi_{E,z}(\omega)|$ and $\ImchiS{zz}$ compared to the previous two cases due to the less solitonic feature of the spin texture. 
In contrast, the additional peaks in the FFM phase are much sharper than the previous ones.

%%%%%%%%%%%%%%%%%%%%%%%%%%%%%%%%%%%%%%%%%%%%%%%%%%%%%%%%%%%%%%%%%%%%%%%%%%

\subsection{Real-space behavior \label{sec:OBC.3}}

\begin{figure}[tb]
\centering
\includegraphics[width=1.0\columnwidth]{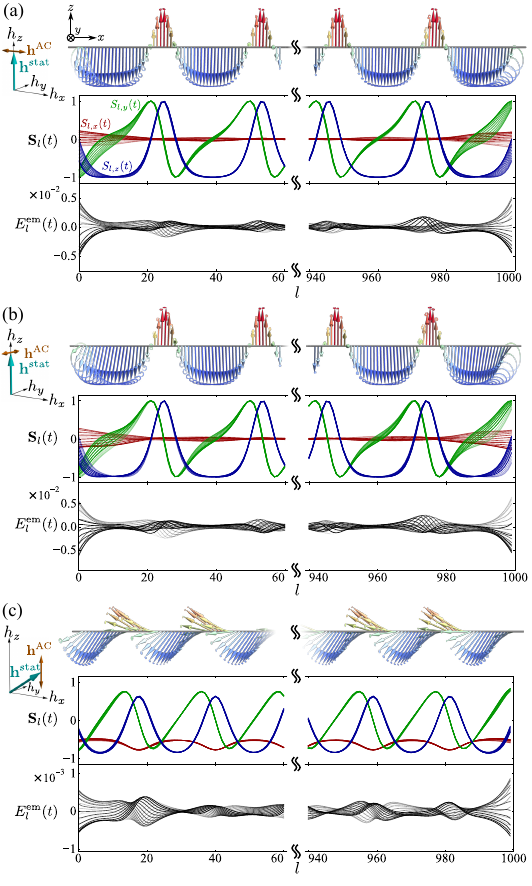}
\caption{
\label{fig:obc_real}
Real-space spin configurations (top and middle) and real-space distributions of the EEF (bottom) in the steady states under the AC magnetic field ${\bf h}^{\rm AC}$. 
(a), (b), and (c) are for ${\bf h}^{\rm AC}\parallel\hat{\bf x}$, ${\bf h}^{\rm AC}\parallel\hat{\bf y}$, and ${\bf h}^{\rm AC}\parallel\hat{\bf z}$, respectively, at 
(a) $(\hst_x, \hst_z)=(0,0.06)$ and $\omega=0.077$, 
(b) $(\hst_x, \hst_z)=(0,0.06)$ and $\omega=0.077$, and 
(c) $(\hst_x, \hst_z)=(0.06,0.04)$ and $\omega=0.061$. 
All the frequencies are set at the values for the edge mode at $\omega^{\rm edge}$; 
the values are denoted by the orange circles in Figs.~\ref{fig:obc_chimax}(d)--\ref{fig:obc_chimax}(f). 
The results are shown for each of the $60$ sites at the left and right edges of the $1000$-site system with the OBC.
The notations are common to those in Fig.~\ref{fig:pbc_real}. 
}
\end{figure}

\begin{figure*}[tb]
\centering
\includegraphics[width=1.0\textwidth]{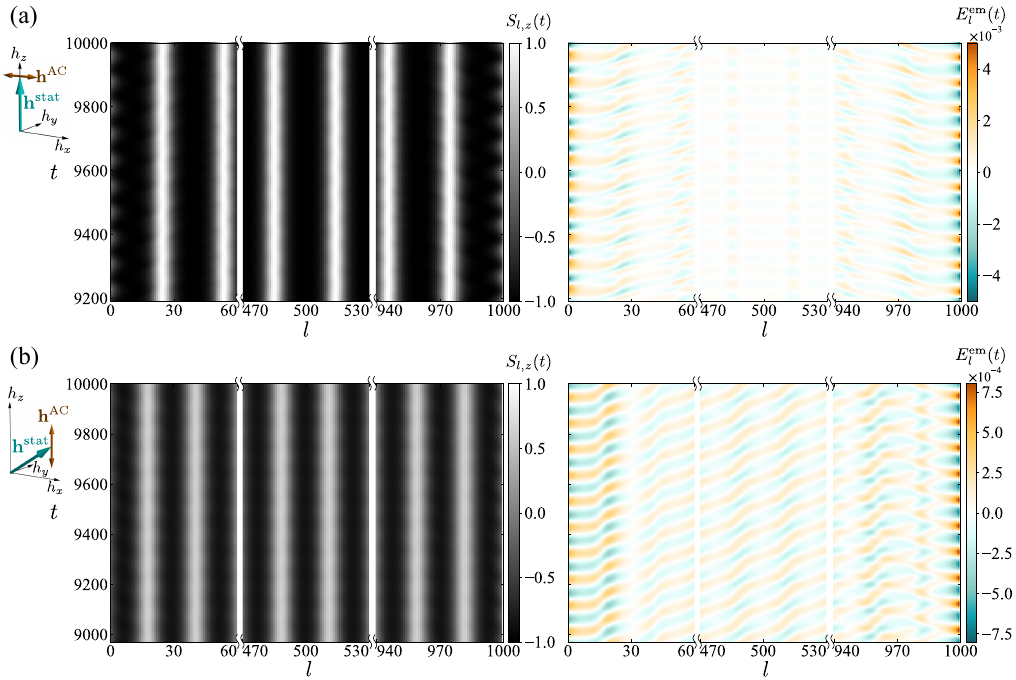}
\caption{
\label{fig:obc_xt}
Real-space and real-time distribution of the $z$ component of spins $S_{l,z}(t)$ (left) and $E^{\rm em}_l(t)$ (right) 
within 10 periods of ${\bf h}^{\rm AC}$ on the plane of $l$ and $t$: 
(a) and (b) correspond to Figs.~\ref{fig:obc_real}(a) and \ref{fig:obc_real}(c), respectively.
The results are shown for 60 sites at the left edge, center, and right edge of the 1000-site system.  
}
\end{figure*}

The spatiotemporal profiles of ${\bf S}_{l}(t)$ and $E^{\rm em}_l(t)$ for the additional mode at $\omega^{\rm edge}$ are shown in Fig.~\ref{fig:obc_real}. 
Here, we display the spin textures and the EEF for each of the 60 sites at the left and right edges of the system with $L=1000$. 
Figure~\ref{fig:obc_real}(a) shows the result for $\hacv\parallel\hatv{x}$ in the CSL state with $(\hst_x, \hst_z)=(0, 0.06)$ and $\omega=0.077$, indicated by the orange circle in Fig.~\ref{fig:obc_chimax}(d). 
In this case, as shown in the top and the middle panels of Fig.~\ref{fig:obc_real}(a), we find that the spin dynamics with $\omega^{\rm edge}$ is localized around the edges, where the spins are twisted in the ground state, known as the chiral surface twist~\cite{Meynell2014}. 
Hence, the additional mode found in Sec.~\ref{sec:OBC.1} is the edge mode. 
Note that the EEF $E_l^{\rm em}(t)$ at both edges has the same sign and its amplitude 
is about ten times larger than that in the bulk, leading to a comparably large contribution to the bulk responses shown in Fig.~\ref{fig:obc_spectra}(a). 
Figure~\ref{fig:obc_xt}(a) shows the time evolution of $S_{l, z}(t)$ and $E^{\rm em}_l(t)$. 
Both patterns obey twofold rotational symmetry about the $z$ axis with time translation by $\frac{\pi}{\omega}$ with respect to the center of the system, but $E^{\rm em}_l(t)$ appears to propagate from the edges to the inside as shown in the right panel, while $S_{l, z}(t)$ on the left does not clearly show such a behavior. 

We observe similar spatiotemporal profiles of the edge mode for $\hacv\parallel\hatv{y}$, as shown in Fig.~\ref{fig:obc_real}(b).  
Meanwhile, as already discussed in Sec.~\ref{sec:OBC.2}, the edge mode for $\hacv\parallel\hatv{z}$ is less significant as shown in Fig.~\ref{fig:obc_real}(c). 
In this case, the EEF propagates from left to right similar to the PBC case, even near the edges, as shown in the right panel of Fig.~\ref{fig:obc_xt}(b), while the pattern near the right edge looks complicated.

\section{Discussion \label{sec:discussion}}

\subsection{Bulk contribution \label{sec:discussion.1}}

Through the calculations for the PBC in Sec.~\ref{sec:PBC}, we showed that the EEF is enhanced at the magnetic resonance. 
An interesting finding is that the EEF resonance peaks are more clearly visible compared to the magnetic ones for higher frequency modes. 
Thus, the measurement of the EEF could be a good probe of the high-frequency magnetic resonances. 
In addition, the EEF is amplified by the solitonic feature of the spin textures which can be controlled by the static magnetic field as well as the direction of the AC magnetic field. 
In particular, it is strongly enhanced when $\hstv$, $\hacv$, and the chiral axis are orthogonal to each other, as exemplified in Figs.~\ref{fig:pbc_chimax}(b) and \ref{fig:pbc_chimax}(e). 
As the total voltage generated by the EEF is proportional to the system size, the solitonic noncollinear spin structure is a promising platform for exploring the emergent electric phenomena, compared to magnetic domain walls~\cite{Yang2009}.

Besides the total EEF, the spatiotemporal profile of the EEF in Sec.~\ref{sec:PBC.3} is important for electronic transport phenomena. 
When $\hst_x=0$, the EEF behaves like a standing wave, and hence, only an AC electric current is expected for electrons coupled to the spin texture. 
When $\hst_x$ becomes nonzero, however, the EEF begins to propagate in the $x$ direction along with the magnon propagation, in addition to the slower drift to the opposite direction 
due to the Archimedean screw dynamics. 
These motions of the EEF give rise to a DC electric current as well as the AC one, 
and hence, the noncollinear spin texture works as a generator of the AC and DC electric currents through the EEF, 
which can be tuned by the static and AC magnetic fields.  
The current direction as well as the velocity may depend on how the electrons couple to the spin texture.

\subsection{Edge contribution \label{sec:discussion.2}}

From the calculations for the OBC in Sec.~\ref{sec:OBC}, we showed that there is a sizable contribution to the EEF from the edges of the system, which can be comparable to or larger than the bulk one depending on the system size. 
It is worth noting that the edge resonances of the EEF are more conspicuous than the magnetic ones. 
Thus, the EEF is also a good probe of the edge modes, in addition to the higher-frequency bulk modes.

In CrNb$_3$S$_6$, which is one of the candidate materials for the monoaxial chiral helimagnets, the lattice spacing between neighboring spins along the chiral axis is about $0.6$~nm and the magnetic modulation period of the CSL phase ranges from $50$ to $100$~nm depending on the magnetic field~\cite{Togawa2012}. 
The situation is similar to that in our calculations, suggesting that a large edge contribution of the EEF comparable to the bulk one is expected for this material even in micro-meter samples. 
We note that while the magnetic resonances were measured for CrNb$_3$S$_6$, the edge contribution has not been clearly observed~\cite{Shimamoto2022}.  
It is thus worth measuring the EEF to identify the edge contributions. 
For this purpose, measurements for samples with different sizes would be useful since the edge contribution is independent of the system size in contrast to the bulk one proportional to the system size.

\subsection{Order estimate \label{sec:discussion.3}}

Finally, let us estimate the amplitude of the EEF expected in real materials.
Experimentally, the  noncollinear spin phase including the CSL has been observed, e.g., in CrNb$_3$S$_6$~\cite{Togawa2012}, MnNb$_3$S$_6$~\cite{Karna2019}, strained Cu$_2$OSeO$_3$~\cite{Okamura2017}, and Yb(Ni$_{1-x}$Cu$_x$)$_3$Al$_9$~\cite{Matsumura2017}. 
The actual value of the EEF depends on the energy scale of the magnetic interactions and the length scale set by the lattice constant in each material. 
By assuming the typical values of the Heisenberg interaction $J$ and the lattice constant being $1$~meV and $1$~nm, respectively, the time unit $t=1$, the frequency $\omega=1$, the magnetic field $h=1$, and the EEF $\bar{E}^{\rm em}=1$ correspond to $\sim 0.66$~psec, $\sim 241$~GHz, $\sim 8.6$~T, and $\sim5\times10^5$~V/m, respectively. 
In our results, we typically obtained $|\chi_{E,\mu}(\omega)| \sim0.01$ at the lowest-energy resonance frequency $\omega_1^{\rm bulk} \sim 0.1 \sim 24.1$~GHz, and hence, the AC magnetic field with $|\hacv|\sim0.01 \sim 860$~Oe can generate the total EEF of $\bar{E}^{\rm em}\sim50$~V/m. 
This leads to the electric voltage $\sim 500$~$\mu$V for a sample with $10$~$\mu$m length scale, which could be experimentally measurable. 
Note that the resonance frequency observed in CrNb$_3$S$_6$ is in good agreement with our results~\cite{Shimamoto2022}. 

Meanwhile, our spatiotemporal profile of the EEF revealed that the local value of $E_l^{\rm em}(t)$ reaches $\sim10^{-2}$, which corresponds to the electric field of $\sim 5$~KV/m. 
Thus, even though the total EEF is relatively small, electrons coupled to the noncollinear spin texture would be largely influenced by such a strong local electric field. 
This may bring about intriguing electric responses, such as the AC and DC electric currents mentioned in Sec.~\ref{sec:discussion.1}. 
Note that the local EEF takes such a large value even in the helical and the conical states where the solitonic feature is absent. 

We note that the generated EEF depends on the value of the Gilbert damping $\alpha$ in Eq.~\eqref{eq:LLG}. 
The EEF associated with the resonance dynamics is roughly proportional to $1/\alpha$. 
This is because the peak intensity of $\chi_{E,\mu}(\omega)$ is approximately proportional to $\chi_{S,\mu\mu}(\omega)$ [see Eq.~\eqref{eq:app_chiEres_h=0}] and $\ImchiS{\mu\mu}$ is proportional to $1/\alpha$. 
Meanwhile, the EEF by the coherent sliding dynamics, which gives the $\omega$-independent contribution, is proportional to $\alpha$, as implied by Eq.~\eqref{eq:chiinf} for small $\alpha$. 
The value of $\alpha$ depends on materials and the realistic estimate is not an easy task, while we used $\alpha=0.04$ as a typical value for ferromagnetic metals~\cite{Mizukami2001, Mizukami2010,Oogane2006,Oogane2010}.

\section{Summary \label{sec:summary}}

To summarize, we have theoretically studied the EEF in a one-dimensional chiral magnet by using numerical simulations based on the LLG equation. 
In the system with the PBC, we clarified that the bulk contribution of $|\chi_{E,\mu}(\omega)|$ consists of the resonance peaks reflecting the magnetic resonances and the $\omega$-independent contribution arising from the coherent sliding dynamics. 
We showed that the peak height of $|\chi_{E,x}(\omega)|$ increases with the solitonic feature of the spin textures; 
it is maximally enhanced when the static magnetic field, the AC one, and the chiral axis are perpendicular to each other. 
Meanwhile, the $\omega$-independent contribution decreases with the vector spin chirality in the direction of the AC magnetic field; it vanishes when the AC magnetic field is perpendicular to the chiral axis. 
Comparing the spectra of $|\chi_{E,\mu}(\omega)|$ with those of $\ImchiS{\mu\mu}$, we revealed that the higher-$\omega$ resonance peaks are more clearly visible in $|\chi_{E,\mu}(\omega)|$ rather than in $\ImchiS{\mu\mu}$, 
since the EEF includes an additional $\omega$-linear factor coming from the time derivative of the spin configuration. 
In addition to the bulk averaged responses, by directly investigating the spatiotemporal profiles of the EEF and the spin texture, 
we revealed that the local EEF associated with the resonance dynamics can take a considerable value even when the total EEF vanishes. 
The amplitude of the EEF is maximized around the solitons where the vector spin chirality is maximized. 
Furthermore, we showed that the EEF behaves like a standing wave when the static magnetic field is perpendicular to the chiral axis. 
In contrast, when the static field has a nonzero component along the chiral axis, the EEF 
is driven to the field direction by the magnon propagation, in addition to the slower drift to the opposite direction due to 
the Archimedean screw dynamics of the spin texture. 
This works as a generator of AC and DC electric currents. 

In addition to the bulk contribution, we have studied the effect of edges of the system under the OBC. 
We found the additional resonance mode in the EEF spectra at a lower frequency, corresponding to the edge mode in the magnetic excitations. 
The frequency of the additional edge mode increases with the solitonic feature of the spin texture, in contrast to the bulk resonance frequencies. 
We also found that the peak intensity of the edge mode can be greater than those of the bulk modes; 
the former is almost system size independent, while the latter is inversely proportional to the system size. 
From the spatiotemporal profiles for the edge mode, we found that the local EEF near the edges can be more than one order of magnitude larger than deep inside the bulk. 
Furthermore, we found that the EEF can be generated even in the FFM phase in contrast to the PBC case. 

Our findings unveil the systematic changes of the EEF for the static and AC magnetic fields in a one-dimensional chiral magnet. 
They would pave the way to enhance the EEF and explore intriguing electronic and magnetic functionalities in chiral magnets. 
To reveal such functionalities explicitly, further studies including electrons coupled to the chiral magnetic textures are desired. 
While the present study was limited to the one-dimensional case, various topological spin textures have been found in two- and three-dimensional magnets, e.g., skyrmions~\cite{Bogdanov1989, Bogdanov1994, Bogdanov1995, Roessler2006}, Bloch points~\cite{Feldtkeller1965, Doring1968, Kotiuga1989} (or equivalently,  magnetic hedgehogs~\cite{Volovik1987, Kanazawa2016, Fujishiro2019}), and hopfions~\cite{Sutcliffe2018,Kent2021,Rybakov2022}. 
It is also intriguing to extend our study to the emergent electric phenomena in these topologically-nontrivial spin textures.

\begin{acknowledgments}
The authors thank J. Kishine, M. Mochizuki, Y. Shimamoto, and Y. Togawa for fruitful discussions.
This research was supported by Grant-in-Aid for Scientific Research Grants (Nos. JP18K03447, JP19H05822, JP19H05825, JP21J20812, and No. JP22K13998), JST CREST (Nos. JP-MJCR18T2 and JP-MJCR19T3), and the Chirality Research Center in Hiroshima University and JSPS Core-to-Core Program, Advanced Research Networks. K.S. was supported by the Program for Leading Graduate Schools (MERIT-WINGS). Parts of the numerical calculations were performed in the supercomputing systems in ISSP, the University of Tokyo.
\end{acknowledgments}

\appendix

\section{$\chi_{E,\mu}(\omega)$ on resonance \label{appendix:chiE_resonance}}

In this Appendix, we discuss $\chi_{E,\mu}(\omega)$ at resonance frequencies analytically in the continuum limit.  
In continuous space, the EEF at position $x$ and time $t$ in Eq.~\eqref{eq:eefloc} is written as 
\begin{eqnarray}
E^{\rm em}(x, t)=\frac{\partial {\bf S}(x,t)}{\partial t}\cdot{\bf C}^{\rm vc}(x,t),
\end{eqnarray}
where ${\bf C}^{\rm vc}(x,t)$ is the vector spin chirality given by 
\begin{eqnarray}
{\bf C}^{\rm vc}(x,t)={\bf S}(x,t)\times\frac{\partial {\bf S}(x,t)}{\partial x}. 
\end{eqnarray}
The Fourier component of the EEF is obtained as 
\begin{eqnarray}
E^{\rm em}(q, \omega)&=&-i\int dq' \int d\omega'~\omega'\notag \\
&&\qquad {\bf S}(q', \omega')\cdot{\bf C}^{\rm vc}(q-q',\omega-\omega').
\label{eq:app_eeffourier}
\end{eqnarray}
Then, at a resonance frequency, the total EEF, $\bar{E}^{\rm em}(\omega)=E^{\rm em}(q=0,\omega)$, 
is well approximated by 
\begin{eqnarray}
\bar{E}^{\rm em}(\omega)
\simeq
-i\omega\int dq ~
{\bf S}(q,\omega)\cdot{\bf C}^{\rm vc}(-q, \omega=0). 
\end{eqnarray}
For the noncollinear spin textures discussed in the main text, ${\bf C}^{\rm vc}(q, 0)$ is dominated by the $q=mQ$ components, where $m$ is an integer and $Q$ is the ordering wave number. 
Thus, $\chi_{E,\mu}(\omega)$ for the resonance dynamics is approximately given by 
\begin{eqnarray}
\chi_{E,\mu}(\omega)
&\simeq&-i\omega 
\sum_{m} \frac{{\bf S}(mQ, \omega)\cdot{\bf C}^{\rm vc}(-mQ, \omega=0)}{h_{\mu}^{\rm AC}(\omega)}. 
\label{eq:app_chiEres}
\end{eqnarray}

In the helical state with $\hst_x=\hst_z=0$, ${\bf C}^{\rm vc}(q, \omega=0)$ consists of only the $q=0$ component ($m=0$). 
Thus, when $\hst_x$ and $\hst_z$ are small and the modulation from the helical spin texture is weak, Eq.~\eqref{eq:app_chiEres} is further approximated by 
\begin{eqnarray}
\chi_{E,\mu}(\omega)\simeq-i\omega \chi_{S,\nu\mu}(\omega)\bar{C}^{\rm vc}_{\nu}, 
\label{eq:app_chiEres_h=0}
\end{eqnarray}
where $\chi_{S,\nu\mu}(\omega)$ is the dynamical spin susceptibility in Eq.~\eqref{eq:chimag} and $\bar{C}^{\rm vc}_{\nu}$ is the $\nu$ component of the vector spin chirality for the ground state given by 
\begin{eqnarray}
\bar{\bf C}^{\rm vc}=\frac{1}{L}\int dx~{\bf C}^{\rm vc}(x, t=0). 
\label{eq:app_vector_chirality}
\end{eqnarray}
In Eq.~\eqref{eq:app_chiEres_h=0}, $\chi_{E,\mu}(\omega)$ is linear in $\omega$. 
This explains why the resonance peaks with high frequencies appear more clearly in $\chi_{E,\mu}(\omega)$ than in 
$\chi_{S,\mu\nu}(\omega)$, as found in Sec.~\ref{sec:PBC.2}.

\section{$\chi_{E,\mu}(\omega)$ in large-$\omega$ region \label{appendix:chiE_limit}}

In this Appendix, we show that $\chi_{E,\mu}(\omega)$ in the case of large $\omega$ can be understood from the vector spin chirality, as in Eq.~\eqref{eq:chiinf}. 
For simplicity, we consider the continuum limit again. 
By plugging Eq.~\eqref{eq:LLG} into Eq.~\eqref{eq:eefloc}, we obtain 
\begin{eqnarray}
&&E^{\rm em}(x,t) =
\frac{1}{1+\alpha^2}\left[
-{\bf S}(x,t)\times{\bf h}^{\rm eff}(x,t) \right. \notag \\
&& \left. + \alpha {\bf S}(x,t)\times\left({\bf S}(x,t)\times{\bf h}^{\rm eff}(x,t)\right)
\right] 
\cdot \left({\bf S}(x,t)\times\frac{\partial{\bf S}(x,t)}{\partial x}\right) \notag \\
&&
=-\frac{1}{1+\alpha^2}{\bf h}^{\rm eff}(x,t)\cdot
\left(\frac{\partial{\bf S}(x,t)}{\partial x}+\alpha {\bf S}(x,t)\times\frac{\partial {\bf S}(x,t)}{\partial x} \right). \notag \\
\label{eq:app_eefloc1}
\end{eqnarray}
Since ${\bf h}^{\rm eff}(x,t)-{\bf h}(t)$ is always perpendicular to $\frac{\partial{\bf S}(x,t)}{\partial x}$, the total EEF is calculated as
\begin{eqnarray}
\bar{E}^{\rm em}(t) 
&=& \frac{1}{L}\int_0^Ldx~E^{\rm em}(x,t) \notag \\
&=&-\frac{1}{L(1+\alpha^2)}
\left[
{\bf h}(t)\cdot\int_0^L dx~\frac{\partial{\bf S}(x,t)}{\partial x} \right. \notag \\
&& \left. 
+\alpha\int_0^L dx~{\bf h}^{\rm eff}(x,t)\cdot\left({\bf S}(x,t)
\times \frac{\partial {\bf S}(x,t)}{\partial x} \right) \right].
\label{eq:app_eefloc2}
\end{eqnarray}
The first term in the square bracket of Eq.~\eqref{eq:app_eefloc2} vanishes in the PBC case, while it remains in the order of $1/L$ in the OBC case. 
Hence, in the following, we focus on the second term in Eq.~\eqref{eq:app_eefloc2}. 
Note that it vanishes when $\alpha=0$, indicating that the dissipative dynamics is essential for generating the bulk EEF~\cite{Kishine2012}.

For sufficiently large $\omega$, the spins hardly follow the AC magnetic field, and hence, it is a good approximation to replace ${\bf S}(x,t)$ with the ground-state spin configuration at $t=0$, ${\bf S}(x,0)$. 
In this approximation, ${\bf h}^{\rm eff}(x, t)$ can be decomposed as 
\begin{eqnarray}
{\bf h}^{\rm eff}(x, t) = {\bf h}^{\rm AC}(t) + {\bf h}^{\rm eff}(x, 0).  
\label{eq:app_heff_decomposition}
\end{eqnarray}
By substituting Eq.~\eqref{eq:app_heff_decomposition} into Eq.~\eqref{eq:app_eefloc2}, the EEF is given by
\begin{eqnarray}
\bar{E}^{\rm em}(t)
&\simeq&
-\frac{\alpha}{L(1+\alpha^2)}{\bf h}^{\rm AC}(t)\cdot\int_0^Ldx~
{\bf S}(x,0)\times \frac{\partial {\bf S}(x,0)}{\partial x} \notag \\
&=&
-\frac{\alpha}{1+\alpha^2}{\bf h}^{\rm AC}(t)\cdot\bar{\bf C}^{\rm vc}.
\label{eq:E_largew}
\end{eqnarray}
Note that the contribution from ${\bf h}^{\rm eff}(x, 0)$ vanishes as $\bar{E}^{\rm em}(0)=0$ and ${\bf h}^{\rm AC}(0)=0$. 
Equation~\eqref{eq:E_largew} leads to Eq.~\eqref{eq:chiinf}.

\bibliography{ref}

\begin{thebibliography}{68}
\bibitem [{3a_()}]{3a_note}%
  \BibitemOpen
  \href@noop {} {}\bibinfo {note} {We note that weak responses remain for
  $\hacv\perp\hstv$ due to nonlinear contributions.}\BibitemShut {Stop}%
\end{thebibliography}

\end{document}